\documentstyle[psfig,epsf,conf_iap]{article}
\newcommand{\lya}{\mbox{${\rm Ly}\alpha$}}
\newcommand{\etal}{et al.}

\def\plottwo#1#2{\centering \leavevmode
\epsfxsize=.38\columnwidth \epsfbox{#1} \hfil
\epsfxsize=.38\columnwidth \epsfbox{#2}}

\begin{document}
\heading{%
%
The Gaseous Extent of Galaxies and the Origin of \lya\ Absorption
Systems at $z < 1$
%
} 
\par\medskip\noindent
\author{%
Hsiao-Wen Chen${^1}$, Kenneth M. Lanzetta${^1}$, John K. Webb${^2}$, and
Xavier Barcons${^3}$
}
\address{%
Department of Physics and Astronomy, State University of New York at Stony 
Brook, Stony Brook, NY 11794--3800, USA
}
\address{%
School of Physics, University of New South Wales, Sydney 2052, NSW; AUSTRALIA
}
\address{%
Instituto de F\'\i sica de Cantabria (Consejo Superior de
Investigaciones Cient\'\i ficas -- Universidad de Cantabria), Facultad
de Ciencias, 39005 Santander, SPAIN
}

\begin{abstract}
We present initial results of a program to obtain and analyze HST WFPC2 images
of galaxies in fields of HST spectroscopic target QSOs. The goal of the program
is to investigate how the properties of \lya\ absorption systems observed in
the spectra of background QSOs vary with the properties of intervening 
galaxies. We found that \lya\ absorption equivalent width depends strongly on
galaxy impact parameter and galaxy $B$-band luminosity, and that the gaseous
extent of individual galaxies scales with galaxy $B$-band luminosity by 
$r\propto L_B^{0.35\pm0.10}$.
\end{abstract}
\section{Introduction}
  The ``forest'' of \lya\ absorption lines observed in QSO spectra has 
provided a unique probe of gas surrounding intervening galaxies.  By directly 
comparing galaxies and \lya\ absorption systems along common lines of sight, 
Lanzetta \etal\ \cite{Lanzetta} found that there exists a distinct 
anti-correlation between \lya\ absorption equivalent width and galaxy impact 
parameter. But the scatter about the mean relation is substantial.
 
  In order to investigate how the amount of gas encountered along the line of 
sight depends on galaxy properties, we have initiated a program to obtain and 
analyze HST WFPC2 images of galaxies identified in fields of HST spectroscopic 
target QSOs.  We measure structural parameters, angular inclinations and 
orientations, luminosities, and rough morphological types of 87 galaxies in 
10 QSO fields. Here we will discuss the primary results and the implications. 
The full analysis is presented in Chen \etal\ \cite{Chen}. 

\section{Results and Discussions}

  To determine how the incidence and extent of tenuous gas around galaxies
depends on galaxy parameters, we assume a simple power law between absorber 
and galaxy properties, and apply a maximum likelihood analysis for 26 
galaxy--absorber pairs and 7 galaxies which do not produce \lya\ absorption 
lines to within sensitive upper limits. Galaxy--absorber pairs are identified 
based on the cross-correlation function, $\xi(v,\rho)$, in order to distinguish
physical pairs from correlated and random pairs \cite{L2}.
\begin{figure}
\centerline{\vbox{
\plottwo{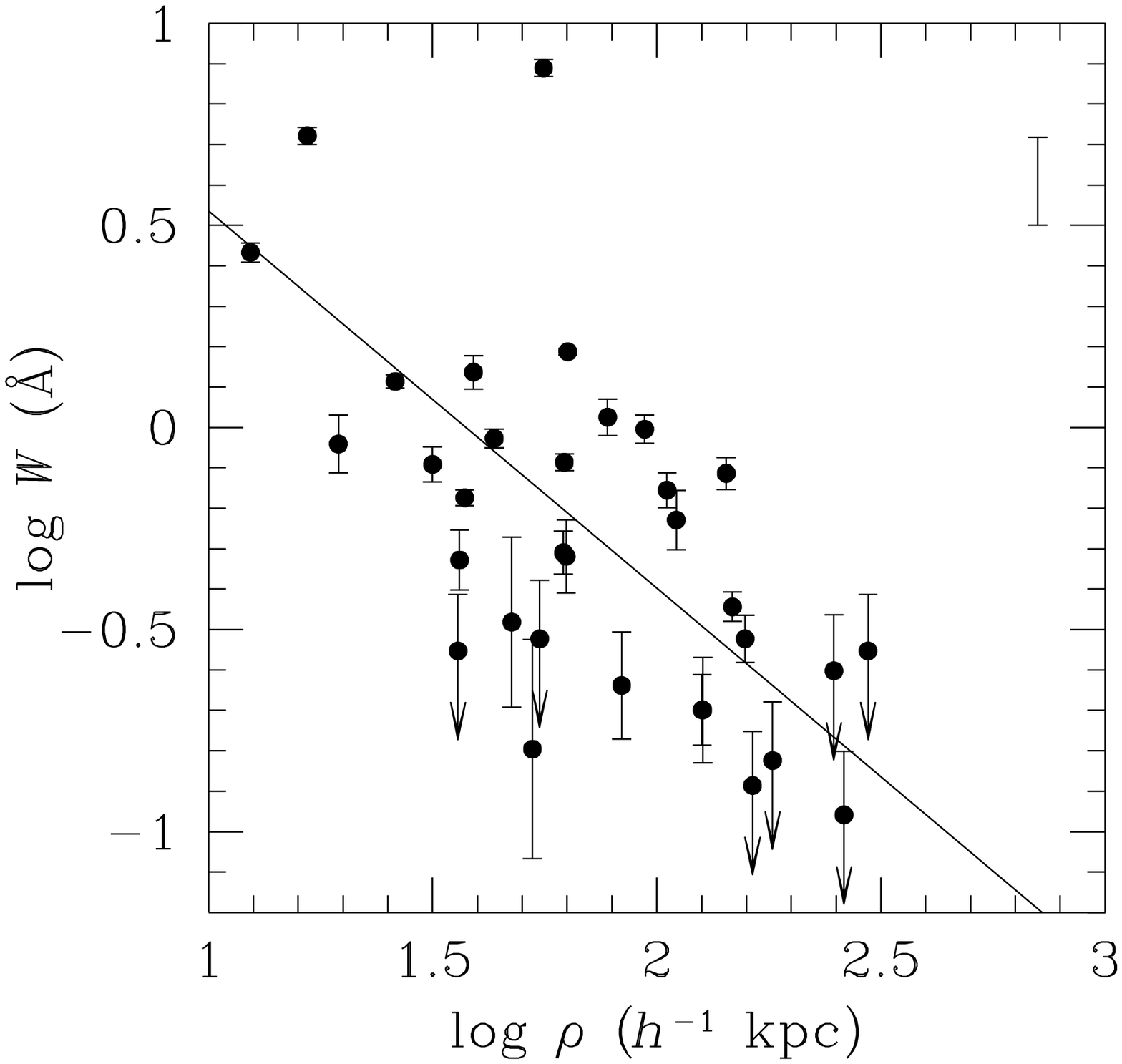}{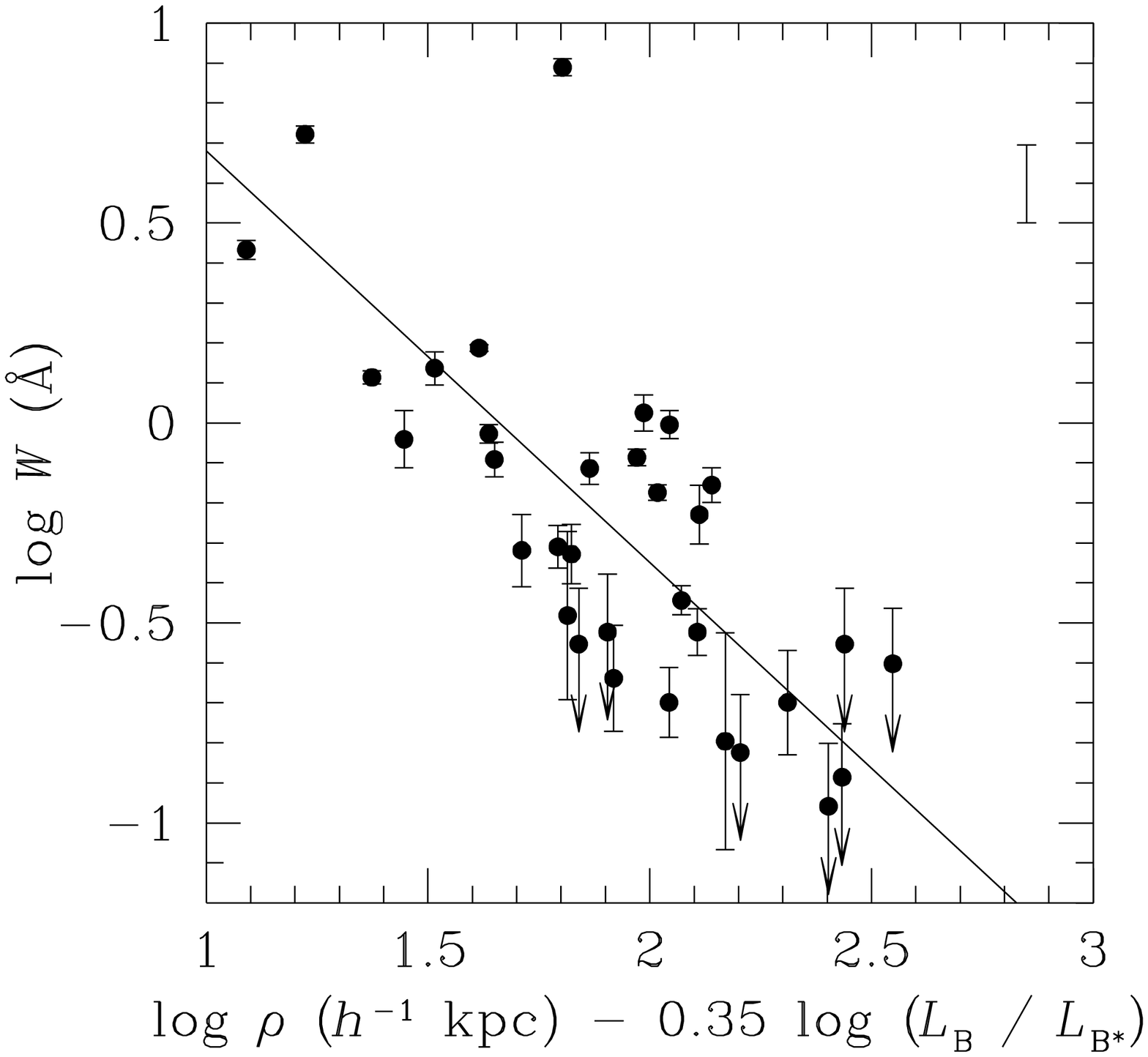}
}}
\caption[]{Distributions of $\log W$ vs. $\log\rho$ and $\log W$ vs. 
$\log\rho-t\log(L_B/{L_B}_*)$. Arrows indicate $3 \sigma$ upper limits. The 
cosmic scatter is shown in the corner.}
\end{figure}

  The primary result of the analysis is that the amount of gas encountered
along the line of sight depends on the galaxy impact parameter and $B$-band
luminosity (Fig 1) but does not depend strongly on the galaxy mean surface
brightness, disk-to-bulge ratio, or redshift. This result {\it confirms} and 
{\it improves} upon the anti-correlation between \lya\ absorption equivalent 
width and galaxy impact parameter found previously by Lanzetta \etal\ 
\cite{Lanzetta}. To summarize: (1) extended gaseous envelopes are a common and 
generic feature of normal galaxies of all types, (2) the gaseous extent of
individual galaxies scales with galaxy luminosity by $r/r_*=(L_B/{L_B}_*)^{t}$
with $t=0.35\pm0.10$ and $r_* = 149\pm 21 \ h^{-1} \ {\rm kpc}$ at $W=0.3$ \AA,
(3) the known gaseous cross sections of known galaxies of luminosity 
$L_B > 0.05 {L_B}_*$ (we adopt a Schechter luminosity function from different 
redshift surveys \cite{Lilly,Ellis,Zucca}) can account for all the \lya\ 
absorption systems with $W>0.3$ \AA\ \cite{Chen}. 
The result strongly suggests that galaxy interactions do not play a role in 
distributing gas around galaxies. In addition, it may provide a means to 
confine the space density of dwarf galaxies.

\acknowledgements{We thank the staff of STScI for their assistance. 
This work was supported by grant AST-9624216 from NSF; grant NAG-53261 from
NASA; and grant GO-0594-80194A, GO-0594-90194A, and GO-0661-20195A from STScI.}


%
\vfill
\end{document}